\documentclass[prb]{revtex4}%
\usepackage{graphicx}
\usepackage{amsmath}
\usepackage{epsfig}
\usepackage{amsfonts}
\usepackage{amssymb}%
\providecommand{\U}[1]{\protect\rule{.1in}{.1in}}
\providecommand{\U}[1]{\protect\rule{.1in}{.1in}}
\begin{document}
\title{\ Graphene-like massless Dirac fermions in Harper systems\\ }
\author{Francisco Claro}
\affiliation{\textit{Facultad de F\'{\i}sica, Pontificia Universidad Cat\'{o}lica de
Chile, Santiago, Chile}}

\begin{abstract}
\noindent It is shown that systems described by Harper's equation
exhibit a Dirac point at the center of the spectrum whenever the
field parameter is a fraction of even denominator. The Dirac point
is formed by the touching of two subbands, and the physics around
such point is characterized by the relative field only, as if the
latter were null at the reference value. Such behavior is consistent
with the nesting property conjectured by Hofstadter, and its
experimental verification would give support to such hypothesis as
well as the Peierls-Onsager ansatz used to arrive at Harper's
equation when crystalline electrons move in a uniform external
magnetic field.\ 

\bigskip


\ \bigskip

\end{abstract}
\maketitle



The relative simplicity with which graphene - carbon single layer
sheets - may be made and handled in the laboratory has drawn much
attention to the physics of massless Dirac particles.\cite{novos2}
In this material electrons moving in two dimensions (2D) near the
Fermi level are subject to an effective energy dispersion law
proportional to momentum rather than the usual momentum squared. The
dynamics is similar to that of photons and phonons, except that in
graphene the particles are charged fermions. They interact among
themselves through the Coulomb force and with external electric and
magnetic fields, which makes them amenable to a varied palette of
experimental manipulation and possible applications.

In graphene carbon atoms are arranged in a two dimensional hexagonal
lattice, obtained as the overlap of two identical triangular
lattices displaced one respect to the other. As a result of the
lattice symmetry the valence and conduction bands have two
inequivalent degenerate points about which the dispersion is linear,
the so called Dirac points.\cite{mclur} The effect is absent in the
square lattice. In this letter we show that for certain values of a
perpendicular magnetic field the square lattice also supports a
Dirac point where two subbands meet, however. Such special values
are defined by a magnetic flux through the unit cell representing a
fraction of flux quanta of even denominator. Furthermore, in the
neighborhood of every one of these values Landau levels emerge from
the Dirac point as they do from zero magnetic field in the hexagonal
graphene lattice. At the reference field the massless Dirac
particles behave as if there was no magnetic field and in its
neighborhood, they appear to respond only to the difference field
much as composite fermions do at major Landau level filling
fractions of even denominator.\cite{jain} These results are a
property of Harper's equation and therefore generic to all systems
governed by such relation. Besides electrons in a square lattice and
a perpendicular magnetic field the equation has appeared in several
contexts, including the quantum Hall effect,\cite{thoul,cla0}
superconducting networks,\cite{panne} nonperiodic solids\cite{sokol}
and electrons in superlattices.\cite{gerha}

The dynamics of Bloch electrons in a square lattice of lattice
constant $a$ and a perpendicular magnetic field $B$ is governed by
Harpers equation,\cite{harper}
\begin{equation}
f_{n+1}+f_{n-1}+2\cos(2\pi\phi n+\nu)f_n=\varepsilon f_n,
\label{harper}
\end{equation}
where $f_n$ is the amplitude of a Wannier state localized at site
$na$ along the x-axis, $n$ an integer, $\phi$ is the magnetic flux
traversing a plaquette in units of the flux quantum $hc/e$, $\nu =
k_ya$ is the dimensionless momentum variable along the y-axis, and
$\varepsilon$ is the energy in units of the hopping integral $t$.
The usual Landau gauge $\mathbf{A} = B(0,x,0)$ is assumed. When the
flux parameter is a rational $\phi = p/q$, $p$ and $q$ integers
prime to each other, then the potential in equations (\ref{harper})
has period $q$ and the set may be closed by selecting solutions with
the property $f_{n+q} = exp(iq\mu)f_n$, $\mu = k_xa$. The condition
for existence of solutions of the resulting $q$ equations for the
unknowns $f_1, f_2, ... f_q$ is that the determinant of the
coefficients,
\begin{equation}
D(\varepsilon,\mu,\nu)=P_q(\varepsilon)-2(\cos q\nu+\cos
q\mu)+(-1)^{\frac{q}{2}}4, \label{det}
\end{equation}
vanishes.\cite{langb} Here $P_q(\varepsilon)$ is a polynomial of
even parity and degree q in $\varepsilon$, independent of the
momentum variables $\nu$ and $\mu$, and with the property
$P_q(0)=0$. The integer $q$ has been assumed to be even. For each
value of $\mu$ and $\nu$ Eq. \ref{det}) has $q$ zeroes which, as
these quantum numbers cover their range, span the $q$ subbands
present in the spectrum. If $q = 4s$, $s$ a positive integer, one
can verify that $\varepsilon = 0, \nu = \mu = 0$ is a solution of
(\ref{det}). Likewise, $\varepsilon = 0$, $\nu = \pm\pi/q$, $\mu =
\pm\pi/q$ are solutions when $q = 2s$, $s$ odd. Zero energy, the
center of the field free band, is thus always in the spectrum. It
corresponds to the edges of two separate subbands that meet at a
single critical point in the Brillouin zone, its center if $q/2$ is
even, and the four equivalent corners if odd. At this point two
neighboring bands meet, never overlapping.\cite{cla1}

That the dispersion near the center of the spectrum is linear in the
magnitude of the momentum follows from the property
$P_q(-\varepsilon)=P_q(\varepsilon)$ for all q even. Near zero
energy one has $P_q(\varepsilon) \approx
-(-1)^{\frac{q}{2}}A(p,q)\varepsilon^2$, where $A(1,2) = 1$, $A(1,4)
= 8$, $A(1,6) = 24$, $A(1,8) = 96-32\sqrt{2}$, $A(3,8) =
96+32\sqrt{2}$ ... are all positive constants.\cite{hase} The
condition on the determinant (\ref{det}) then simplifies to
\begin{equation}
A(p,q)\varepsilon^2+2(-1)^{\frac{q}{2}}(\cos q\nu+\cos q\mu)-4 = 0.
\label{condition}
\end{equation}
For $q/2$ even this gives near the Brillouin zone center
\begin{equation}\label{disp}
\varepsilon = \pm C(p,q)\sqrt{\nu^2+\mu^2},
\end{equation}
where $C(p,q)=qA(p,q)^{-1/2}$ is a magnetic field dependent velocity
in units $ta/\hbar$. A similar relation is obtained near each
Brillouin zone corner for $q/2$ odd, with the momentum measured
respect to the appropriate zone corner. Figure 1(a) shows a Dirac
point placed at the zone corners corresponding to $q = 2$, in which
case form (\ref{condition}) is exact. Figure 1(b) is for $q = 4$ and
exemplifies a Dirac point placed at the center of the zone. Specular
reflection over the momentum plane gives the dispersion for negative
energy, corresponding to holes. We note that although a 2D hexagonal
crystal exhibits two inequivalent Dirac points at zero magnetic
field, they are lost when the field is turned on except at special
values of the flux of the form $\phi = n \pm 1/6$, n an integer or
zero, where a single such critical point occurs.\cite{cla2}

\begin{figure}[h]
\includegraphics[width=5in]{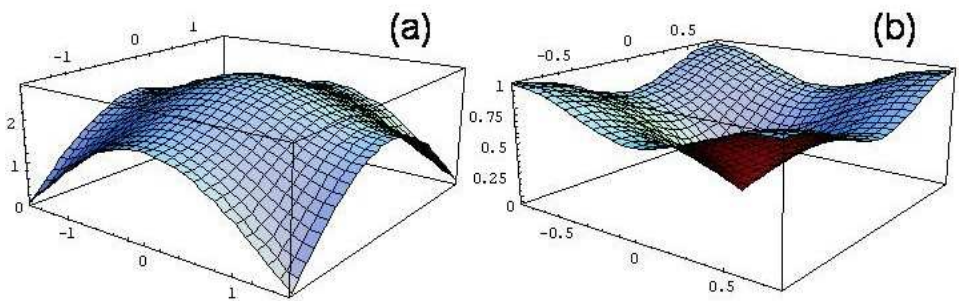}
\caption{Energy dispersion for
(a) $q =2$ and (b) $q = 4$. The energy (vertical axis) is in units
of the band parameter t, while the perpendicular plane represents
dimensionless momentum.}%
\label{disp1}%
\end{figure}

As shown by Hofstadter, the spectrum of Harpers equation has a
recursive structure of nonoverlapping subbands arranged in such ways
that, displayed over the fundamental cell $0\leq \phi\leq 1$,
resemble a butterfly with open wings.\cite{hofs} He conjectured a
nesting property that makes each subband in the spectrum a replica
at some recursive level of the field-free band, about which a
geometrically distorted version of the whole graph develops. One
consequence is that, for instance, at $\phi^{\prime} = \phi + \delta
\phi$ a cyclotron frequency $\omega = q\delta \phi/\hbar
g(\phi,\varepsilon)$ may be defined in the neighborhood of any
subband pertaining to the spectrum at flux $\phi$, where
$g(\phi,\varepsilon)$ is the density of states at energy
$\varepsilon$. Note that this expression scales as $\delta \phi$, as
expected from the notion that the subband in question is a recursive
replica of the field free parent band at $\phi = 0$.\cite{cla3}

To study the special situation when two subbands touch at energy
$\varepsilon = 0$ we take advantage of the effective hamiltonian
formalism.\cite{taut} In this theory, the original problem of an
electron moving in the presence of a square lattice potential and an
external magnetic field $B'$ in the neighborhood of a given subband
belonging to the spectrum at field $B$, is replaced by the
hamiltonian problem defined by
\begin{equation}\label{taut1}
    H_r = \varepsilon_r
    (\frac{1}{\hbar}[\textbf{p}+\frac{e}{c}\Delta\textbf{A}(\textbf{r})],\phi),
\end{equation}
where $\varepsilon_r(\textbf{k},\phi)$ is the dispersion law in
subband $r$ and $\Delta\textbf{A}(\textbf{r})= \delta B(0,x,0)$,
with $\delta B = B^{\prime}-B$. If the integer $r=1,2,...,q$ labels
the subbands in order of increasing energy then $\varepsilon_{q/2}$,
$\varepsilon_{q/2+1}$ touch at zero energy, near which the
dispersion obeys Eq. (\ref{condition}). In that neighborhood the
dispersion is linear and the usual methods of quantum mechanics can
be employed to obtain the associated spectrum and
eigenfunctions.\cite{seme} The former is found to have the form
\begin{equation}\label{taut2}
E_{n}=sgn(n)2qt \sqrt{\frac{\pi}{A(p,q)}|n\delta\phi|},
\end{equation}
where $\delta\phi = e\delta B a^2/hc$ is the flux traversing a unit
cell, measured with respect to the reference value $\phi = p/q$, and
$n=0,\pm1,\pm2,...$ is a Landau level index. Figure 2 shows this
expression evaluated in the neighborhood of flux $1/2$ up to $n = 4$
(solid lines), together with the associated spectrum given by roots
of Eq. (\ref{det}) at a few rational values of the flux in that
neighborhood (dots). In the latter the Landau levels have a width
and possibly internal structure, only that so narrow that it is not
resolved in the scale of the figure. The agreement is excellent at
low Landau levels, though it deteriorates slowly as the Landau index
increases and the relative flux grows. The figure shows the positive
quadrant only and it repeats for negative flux and negative energy,
specularly reflected over the proper axes.

\begin{figure}[h]
\includegraphics[width=3.0in]{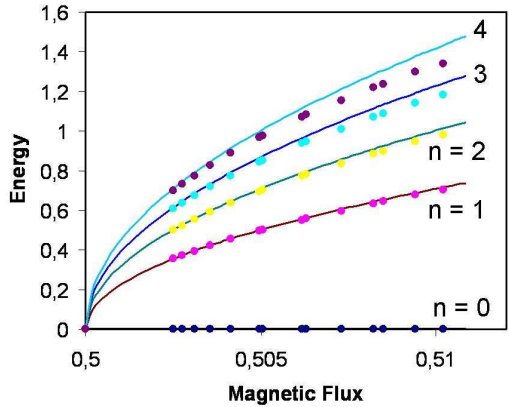}
\caption{Landau levels emerging
from the Dirac point at 0.5 flux quanta per unit cell (solid lines),
and
solutions to Harper's equation in the same neighborhood (dots).}%
\label{energy}%
\end{figure}

The number of states in each Landau level may be obtained from the
gap labeling theorem, according to which the statistical weight of
states below any gap in the spectrum is given by a continuous
function of the field, of the form $W = M\phi + N$, with $M, N$
integers.\cite{wann} The value $\varepsilon = 0$ divides the
spectrum in two specularly symmetric halves, so at that energy $W =
1/2$. Using these two facts one readily finds that the number of
states of any level in the Landau fan emerging from the Dirac point
at flux $\phi = p/q$, $q$ even, is simply $D = q\delta\phi$. For
instance, for $\phi = 1/2$, the number of states per cell below a
gap reaching the apex at $\varepsilon = 0$ has the form $W_{N} =
(1+2N)\phi - N$, with $N$ an integer. The number of states between
two neighboring gaps is then $W_{N+1}- W_{N} = 2\phi - 1 =
2\delta\phi$, in accordance with the general result just described.
The number of states grows linearly with the relative flux as it
does for free 2D electrons in a magnetic field, yet the total number
is $q$ times larger. Inclusion of spin degeneracy is achieved
through an additional factor of 2.

Landau levels are separated by sizable gaps traversing the energy
versus field graph and meeting at the Dirac point. Here the gap
labeling theorem yields the equation $s = Mp + 2sN$, $s = q/2$,
which restricts the values of M to the set $M = s(2n + 1)$, $n$ an
integer. Since $M$ may be identified with the dimensionless
conductance,\cite{thoul} in the neighborhood of a Dirac point at
flux $p/q$ one expects the inverse Hall resistance to be quantized
according to
\begin{equation}\label{conduct}
R_{xy}^{-1} = q(n+\frac{1}{2})\frac{e^{2}}{h}.
\end{equation}
Because q is even, this quantity will always involve integer
multiples of the quantum of conductance $e^{2}/h$ even when the spin
degeneracy is fully resolved. In the simple case $q = 2$ the
sequence of multiples for electrons in the latter case is 1, 3, 5,
... but if the Zeeman energy is small the sequence is the same as
that observed in graphene, e. g. 2, 6, 10, ....

The existence of Dirac points in Harper systems as shown, could play
an important role in settling questions concerning such systems. One
issue is the correctness of the Peierls-Onsager substitution method
used in deriving Harpers equation from a tight binding field-free
band, a yet unproven ansatz.\cite{peier,onsag} It works well in the
semiclassical regime at small fields but its performance when the
flux per cell approaches one flux quanta is largely unknown. Another
issue is the nesting hypothesis used by Hofstadter to describe the
recursivity of the spectrum, mentioned above.\cite{hofs} Full
experimental verification of the complex subband spectrum and
associated dynamics predicted by Harper's equation would give strong
support to both the Peierls-Onsager ansatz and the nesting
hypothesis.

In order to enter the desired region of about one flux quantum per
unit cell, experimentally unreachable magnetic fields in the order
of $10^{5}$ Tesla are required by ordinary atomic solids. However,
artificial crystalline potentials with a lattice constant of about
10nm are possible, allowing to work in the few Tesla regime.
Features of the Hofstadter butterfly have already been recognized as
Landau level internal structure in the weak lattice potential
limit.\cite{enssl,vonk} As better samples become available it is
hoped that the search of Shubnikov-de Hass oscillations and quantum
Hall plateaus as one moves away from a Dirac point such as the one
at $\phi = 1/2$ will give decisive information on the correctness of
the Peierls-Onsager ansatz and the nesting hypothesis.

Helpful discussions with Greg Huber are greatly acknowledged. This
research was funded by Fondecyt, Grant 1060650.

\end{document}